\documentclass[a4paper,11pt]{article}
\usepackage{jheppub}

\usepackage{siunitx}
\DeclareSIUnit\barn{b}
\usepackage{booktabs}
\usepackage[italic]{heppennames2}

\usepackage{tikz}
\usetikzlibrary{arrows,shapes,backgrounds,shapes.callouts}
\usepackage[compat=1.1.0]{tikz-feynman}
\tikzfeynmanset{warn luatex=false}
\tikzfeynmanset{ with arrow/.style = { 
   decoration={
     markings,
     mark=at position 0.5 
          with {\arrow[xshift=0.4mm]{Stealth[black,width=1mm,length=1.3mm]}}
     },  
   postaction=decorate}
}
\tikzfeynmanset{
    yourboson/.style={
        decorate,
        decoration={
            snake,
            mirror,
            amplitude=0.6mm,
            segment length=2.16mm,
            pre,
            pre length=0pt,
            post,
            post length=0pt
        }
    }
}

\arxivnumber{2407.03468}

\title{EFT at JADE: a case study}

\author[1]{Jonathan S. Wilson}
\affiliation{Department of Physics,
Baylor University,
One Bear Place \#97316,
Waco, TX 76798-7316}

\emailAdd{Jon\_Wilson2@baylor.edu}

\abstract{
As we use the standard model effective field theory to search for signs of new physics beyond the direct reach of the LHC, we often wonder what we may learn from the effective field theory, and what it would look like to make a discovery via effective field theory.
This article presents a case study that provides some answers to these questions.
We apply the low-energy effective field theory to $\Pep\Pem \to \PGmp\PGmm$ data below the \PZ boson mass from the JADE experiment at DESY.
The low-energy effective field theory allows the observation of physics beyond QED in the JADE data and furthermore, by matching the Wilson coefficients to the electroweak theory, a rough measurement of the masses of the \PW and \PZ bosons is possible.
The ability to make this rough measurement challenges the conventional wisdom that an observation of new physics via EFT tells us nothing about the nature of that new physics.
A measurement of this quality would have been sufficient to guide the construction of colliders such as the super proton-antiproton synchrotron or the large electron-positron collider, and so we anticipate that a discovery of new physics via effective field theory at the LHC would be similarly sufficient to guide the construction of future colliders.
}

\begin{document}
\maketitle
\flushbottom

\section{Introduction}

The LHC today boasts a robust program searching for signs of physics beyond the standard model (SM) using the SM effective field theory (SMEFT)~\cite{Ethier:2021bye}.
The question of what an observation of new physics via SMEFT would teach us is often raised, and the response is generally that, without more targeted data analysis and probably a higher-energy experiment, we will not be able to learn anything about the new physics beyond its existence, not even the energy scale of the new physics.

We have performed a case study that challenges this assumption.
By examining data from below the \PZ boson mass using the low-energy effective field theory (LEFT) and matching the LEFT Wilson coefficients to the electroweak theory, we show that substantial knowledge about the energy scale of new physics can be extracted from nothing more than the measured Wilson coefficients.
This suggests that, in the event of an observation via SMEFT at the LHC, we will similarly be able to obtain substantial knowledge about the nature of the new physics even without a higher-energy experiment, contrary to the conventional wisdom.

This case study uses $\Pep\Pem \to \PGmp\PGmm$ data from the JADE experiment at DESY, and ignores all of the other data relevant to electroweak effects that was available at the time.
It is therefore a counter-historical case study but it more closely mimics a hypothetical future in which indications of physics beyond the SM have been observed at the LHC via SMEFT measurements.

We describe the JADE data in section~\ref{sec:data}, the LEFT and its predictions in section~\ref{sec:LEFT}, the fit to the JADE data to measure the LEFT Wilson coefficients in section~\ref{sec:fit}, the matching of the LEFT Wilson coefficients to the electroweak theory predictions in section~\ref{sec:match}, and the measurement of the masses of the \PW and \PZ bosons in section~\ref{sec:masses}.

\section{Data}
\label{sec:data}

The JADE (JApan, Deutschland, and England) experiment at the PETRA particle accelerator at DESY was a general-purpose particle detector~\cite{NAROSKA198767}.
It recorded $e^+e^-$ collisions from 1979 to 1986, with center-of-mass energies ranging from 12 to \SI{46.6}{\giga\eV}.

JADE measures the differential cross section for electron-positron annihilation to a pair of muons as a function of the angle, in the center-of-mass frame, between the incoming electron and outgoing muon momenta~\cite{JADE_mu_AFB}.
This measurement is performed at four center-of-mass energies: 13.8, 22.0, 34.4, and \SI{42.4}{\giga\eV}.
A forward-backward asymmetry is observed, which increases with center-of-mass energy and is consistent with the predictions of the electroweak theory within roughly $1.3\sigma$.

The differential cross section, multiplied by the Mandelstam variable $s$ and divided by the bin width, is measured in several bins of $\cos \theta$, as shown in tables~\ref{tab:diffXS} and \ref{tab:diffXS2} as well as figure 2 of ref.~\cite{JADE_mu_AFB}.
The binning depends on the center-of-mass energy at which the measurement is performed.

The measured differential cross sections are corrected bin-by-bin for QED contributions to order $\alpha^3$ as reported in ref.~\cite{JADE_mu_AFB}.
These corrections amount to an inclusive forward-backward asymmetry of $+2\%$ for $\left\lvert\cos \theta\right\rvert < 0.8$ and $+6.3\%$ for the two forward bins around $\left\vert\cos \theta\right\rvert = 0.89$ at $\sqrt{s} = \SI{34.4}{\giga\eV}$.
The uncorrected measurement is not available.
Because the measurement has already been corrected for these higher order QED effects, we analyze it using only leading order calculations.

\begin{table}
    \centering
    \begin{tabular}{ccccc}
        \toprule
        Bin & Bin width $w$ & \multicolumn{3}{c}{$\left(s / w\right) \left(\mathrm{d} \sigma / \mathrm{d} \cos \theta\right)$ [\si{\nano\barn\giga\eV\squared}]} \\
         \cmidrule(lr){3-5}
         & & \SI{13.8}{\giga\eV} & \SI{22.0}{\giga\eV} & \SI{42.4}{\giga\eV} \\
        \midrule
        $(          -0.8,          -0.6)$ & 0.2 & \num{6.91( 93)} & \num{6.95(111)} & \num{8.58(106)} \\
        $(          -0.6,          -0.4)$ & 0.2 & \num{6.53( 95)} & \num{7.06(117)} & \num{7.53( 90)} \\
        $(          -0.4,          -0.2)$ & 0.2 & \num{5.00( 76)} & \num{6.99(111)} & \num{7.10( 94)} \\
        $(          -0.2,\phantom{-}0.0)$ & 0.2 & \num{5.23( 80)} & \num{4.29( 88)} & \num{5.37( 83)} \\
        $(\phantom{-}0.0,\phantom{-}0.2)$ & 0.2 & \num{6.30( 85)} & \num{4.68( 95)} & \num{5.57( 83)} \\
        $(\phantom{-}0.2,\phantom{-}0.4)$ & 0.2 & \num{4.92( 79)} & \num{6.87(112)} & \num{4.54( 78)} \\
        $(\phantom{-}0.4,\phantom{-}0.6)$ & 0.2 & \num{7.49( 94)} & \num{5.17( 94)} & \num{7.57( 91)} \\
        $(\phantom{-}0.6,\phantom{-}0.8)$ & 0.2 & \num{7.37( 96)} & \num{5.34( 99)} & \num{4.81( 79)} \\
        \bottomrule
    \end{tabular}
    \caption{The JADE measurement of the differential cross section $\left(s / w\right) \left(\mathrm{d} \sigma / \mathrm{d} \cos \theta\right)$, where $w$ is the bin width, in several bins at center-of-mass energies of 13.8, 22.0, and \SI{42.4}{\giga\eV}.}
    \label{tab:diffXS}
\end{table}

\begin{table}
    \centering
    \begin{tabular}{ccc}
        \toprule
        Bin & Bin width $w$ & $\left(s / w\right) \left(\mathrm{d} \sigma / \mathrm{d} \cos \theta\right)$ [\si{\nano\barn\giga\eV\squared}] \\
        \midrule
        $(          -1.00,           -0.80)$ & 0.20 & \num{9.15(110)} \\
        $(          -0.80,           -0.64)$ & 0.16 & \num{8.56( 47)} \\
        $(          -0.64,           -0.48)$ & 0.16 & \num{7.57( 45)} \\
        $(          -0.48,           -0.32)$ & 0.16 & \num{6.58( 38)} \\
        $(          -0.32,           -0.16)$ & 0.16 & \num{5.62( 33)} \\
        $(          -0.16, \phantom{-}0.00)$ & 0.16 & \num{5.93( 37)} \\
        $(\phantom{-}0.00, \phantom{-}0.16)$ & 0.16 & \num{4.91( 36)} \\
        $(\phantom{-}0.16, \phantom{-}0.32)$ & 0.16 & \num{5.24( 41)} \\
        $(\phantom{-}0.32, \phantom{-}0.48)$ & 0.16 & \num{5.40( 36)} \\
        $(\phantom{-}0.48, \phantom{-}0.64)$ & 0.16 & \num{5.84( 39)} \\
        $(\phantom{-}0.64, \phantom{-}0.80)$ & 0.16 & \num{6.30( 40)} \\
        $(\phantom{-}0.80, \phantom{-}1.00)$ & 0.20 & \num{8.30(104)} \\
        \bottomrule
    \end{tabular}
    \caption{The JADE measurement of the differential cross section $\left(s / w\right) \left(\mathrm{d} \sigma / \mathrm{d} \cos \theta\right)$, where $w$ is the bin width, in several bins at a center-of-mass energy of \SI{34.4}{\giga\eV}.}
    \label{tab:diffXS2}
\end{table}

\section{The low-energy effective field theory}
\label{sec:LEFT}

The low-energy effective field theory (LEFT)~\cite{LEFT} describes physics below the electroweak scale.
The \PW, \PZ, and Higgs bosons and the top quark are integrated out of the SMEFT to obtain the LEFT.
In the most general flavor assumptions, this produces a total of 6083 operators, including dimensions 3, 5, and 6 and allowing CP violation.
Of those, 12 operators affect $\Pep\Pem \to \PGmp\PGmm$ at tree level with a single operator insertion and do not produce CP violation.
These operators, which are listed in table~\ref{tab:operators}, can be categorized as dipole operators, $s$-channel vector operators, $t$-channel vector operators, $s$-channel scalar operators, and $t$-channel scalar operators.
Only the $s$-channel vector operators have nonzero values in the SM in the limit of massless fermions, as indicated in table~\ref{tab:operators}.
Nonetheless, we include all 12 in the analysis that follows.

\begin{table}
    \centering
    \renewcommand{\arraystretch}{1.6}
    \begin{tabular}{cccc}
    \toprule
    Wilson coefficient & Flavor indices & Operator definition & Nonzero in SM \\
    \midrule
    \multicolumn{4}{c}{Dipole operators}\\
    $C_{\substack{\Pe\PGg \\ pr}}$ & $pr = \Pe\Pe$   & $\overline{\Pe}_{Lp} \sigma^{\mu\nu} \Pe_{Rr} F_{\mu \nu} + \text{h.c.}$ & \\
    $C_{\substack{\Pe\PGg \\ pr}}$ & $pr = \PGm\PGm$ & $\overline{\Pe}_{Lp} \sigma^{\mu\nu} \Pe_{Rr} F_{\mu \nu} + \text{h.c.}$ & \\
    \multicolumn{4}{c}{$s$-channel vector operators}\\
    $C^{V,LL}_{{prst}}$ & $prst = \Pe\Pe\PGm\PGm$ & $\left(\overline{\Pe}_{Lp} \gamma^{\mu} \Pe_{Lr}\right) \left(\overline{\Pe}_{Ls} \gamma_{\mu} \Pe_{Lt}\right)$ & * \\
    $C^{V,RR}_{{prst}}$ & $prst = \Pe\Pe\PGm\PGm$ & $\left(\overline{\Pe}_{Rp} \gamma^{\mu} \Pe_{Rr}\right) \left(\overline{\Pe}_{Rs} \gamma_{\mu} \Pe_{Rt}\right)$ & * \\
    $C^{V,LR}_{{prst}}$ & $prst = \Pe\Pe\PGm\PGm$ & $\left(\overline{\Pe}_{Lp} \gamma^{\mu} \Pe_{Lr}\right) \left(\overline{\Pe}_{Rs} \gamma_{\mu} \Pe_{Rt}\right)$ & * \\
    $C^{V,LR}_{{prst}}$ & $prst = \PGm\PGm\Pe\Pe$ & $\left(\overline{\Pe}_{Lp} \gamma^{\mu} \Pe_{Lr}\right) \left(\overline{\Pe}_{Rs} \gamma_{\mu} \Pe_{Rt}\right)$ & * \\
    \multicolumn{4}{c}{$t$-channel vector operators}\\
    $C^{V,LL}_{{prst}}$ & $prst = \Pe\PGm\PGm\Pe$ & $\left(\overline{\Pe}_{Lp} \gamma^{\mu} \Pe_{Lr}\right) \left(\overline{\Pe}_{Ls} \gamma_{\mu} \Pe_{Lt}\right)$ & \\
    $C^{V,RR}_{{prst}}$ & $prst = \Pe\PGm\PGm\Pe$ & $\left(\overline{\Pe}_{Rp} \gamma^{\mu} \Pe_{Rr}\right) \left(\overline{\Pe}_{Rs} \gamma_{\mu} \Pe_{Rt}\right)$ & \\
    $C^{V,LR}_{{prst}}$ & $prst = \Pe\PGm\PGm\Pe$ & $\left(\overline{\Pe}_{Lp} \gamma^{\mu} \Pe_{Lr}\right) \left(\overline{\Pe}_{Rs} \gamma_{\mu} \Pe_{Rt}\right)$ & \\
    $C^{V,LR}_{{prst}}$ & $prst = \PGm\Pe\Pe\PGm$ & $\left(\overline{\Pe}_{Lp} \gamma^{\mu} \Pe_{Lr}\right) \left(\overline{\Pe}_{Rs} \gamma_{\mu} \Pe_{Rt}\right)$ & \\
    \multicolumn{4}{c}{$s$-channel scalar operators}\\
    $C^{S,RR}_{{prst}}$ & $prst = \Pe\Pe\PGm\PGm$ & $\left(\overline{\Pe}_{Lp} \Pe_{Rr}\right) \left(\overline{\Pe}_{Ls} \Pe_{Rt}\right) + \text{h.c.}$ & \\
    \multicolumn{4}{c}{$t$-channel scalar operators}\\
    $C^{S,RR}_{{prst}}$ & $prst = \Pe\PGm\PGm\Pe$ & $\left(\overline{\Pe}_{Lp} \Pe_{Rr}\right) \left(\overline{\Pe}_{Ls} \Pe_{Rt}\right) + \text{h.c.}$ & \\
    \bottomrule
    \end{tabular}
    \caption{The 12 LEFT operators that can affect $\Pep\Pem \to \PGmp\PGmm$ at tree-level with a single operator insertion.  The 4 operators that have nonzero Wilson coefficients in the SM in the limit of massless fermions are marked with an asterisk in the right column.}
    \label{tab:operators}
\end{table}

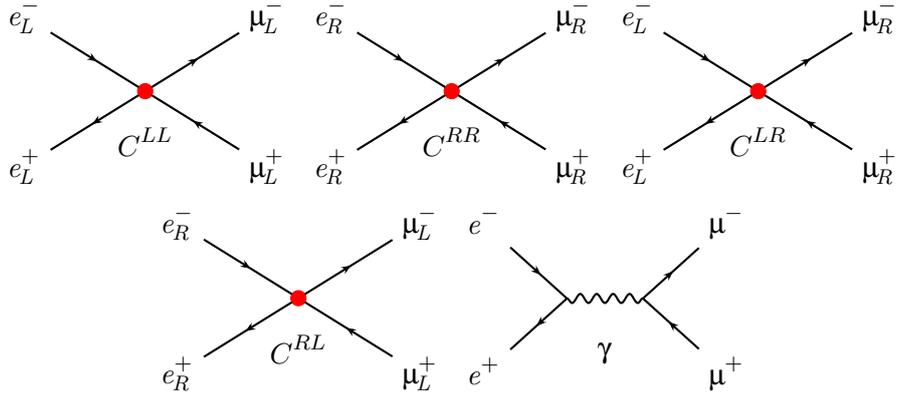
\begin{figure}
    \centering
    \begin{tikzpicture}
        \begin{feynman}[large]
            \vertex (1) at (-1.4, 2.0) {\HepParticle{\Pe}{}{-}};
            \vertex (2) at (-1.4, 0.0) {\HepParticle{\Pe}{}{+}};
            \vertex (A) at (-0.5, 1.0);
            \vertex (B) at (0.5, 1.0);
            \vertex (3) at (1.4, 2.0) {\HepParticle{\PGm}{}{-}};
            \vertex (4) at (1.4, 0.0) {\HepParticle{\PGm}{}{+}};
            \diagram*{
                (1) -- [fermion] (A) -- [fermion] (2),
                (4) -- [fermion] (B) -- [fermion] (3),
                (A) -- [yourboson] (B)
            };
        \end{feynman}
        \node [above] {\PGg};
    \end{tikzpicture}
    \\
    \begin{tikzpicture}
        \begin{feynman}[large]
            \vertex (1) at (-1.4, 2.0) {\HepParticle{\Pe}{}{-}};
            \vertex (2) at (-1.4, 0.0) {\HepParticle{\Pe}{}{+}};
            \vertex[dot,red] (A) at (-0.5, 1.0) {};
            \vertex (B) at (0.5, 1.0);
            \vertex (3) at (1.4, 2.0) {\HepParticle{\PGm}{}{-}};
            \vertex (4) at (1.4, 0.0) {\HepParticle{\PGm}{}{+}};
            \diagram*{
                (1) -- [fermion] (A) -- [fermion] (2),
                (4) -- [fermion] (B) -- [fermion] (3),
                (A) -- [yourboson] (B)
            };
        \end{feynman}
        \node [above] {$C_{\substack{\Pe\PGg \\ \Pe\Pe}}$};
    \end{tikzpicture}
    \begin{tikzpicture}
        \begin{feynman}[large]
            \vertex (1) at (-1.4, 2.0) {\HepParticle{\Pe}{}{-}};
            \vertex (2) at (-1.4, 0.0) {\HepParticle{\Pe}{}{+}};
            \vertex (A) at (-0.5, 1.0);
            \vertex[dot,red] (B) at (0.5, 1.0) {};
            \vertex (3) at (1.4, 2.0) {\HepParticle{\PGm}{}{-}};
            \vertex (4) at (1.4, 0.0) {\HepParticle{\PGm}{}{+}};
            \diagram*{
                (1) -- [fermion] (A) -- [fermion] (2),
                (4) -- [fermion] (B) -- [fermion] (3),
                (A) -- [yourboson] (B)
            };
        \end{feynman}
        \node [above] {$C_{\substack{\Pe\PGg \\ \PGm\PGm}}$};
    \end{tikzpicture}
    \\
    \begin{tikzpicture}
        \begin{feynman}[large]
            \vertex (1) at (-1.4, 2.0) {\HepParticle{\Pe}{L}{-}};
            \vertex (2) at (-1.4, 0.0) {\HepParticle{\Pe}{L}{+}};
            \vertex[dot,red] (A) at (0.0, 1.0) {};
            \vertex (3) at (1.4, 2.0) {\HepParticle{\PGm}{L}{-}};
            \vertex (4) at (1.4, 0.0) {\HepParticle{\PGm}{L}{+}};
            \diagram*{
                (1) -- [fermion] (A) -- [fermion] (2),
                (4) -- [fermion] (A) -- [fermion] (3)
            };
        \end{feynman}
        \node [above=-2ex] {$C^{V,LL}_{{\Pe\Pe\PGm\PGm}}$};
    \end{tikzpicture}
    \begin{tikzpicture}
        \begin{feynman}[large]
            \vertex (1) at (-1.4, 2.0) {\HepParticle{\Pe}{L}{-}};
            \vertex (2) at (-1.4, 0.0) {\HepParticle{\Pe}{L}{+}};
            \vertex[dot,red] (A) at (0.0, 1.0) {};
            \vertex (3) at (1.4, 2.0) {\HepParticle{\PGm}{R}{-}};
            \vertex (4) at (1.4, 0.0) {\HepParticle{\PGm}{R}{+}};
            \diagram*{
                (1) -- [fermion] (A) -- [fermion] (2),
                (4) -- [fermion] (A) -- [fermion] (3)
            };
        \end{feynman}
        \node [above=-2ex] {$C^{V,RR}_{{\Pe\Pe\PGm\PGm}}$};
    \end{tikzpicture}
    \begin{tikzpicture}
        \begin{feynman}[large]
            \vertex (1) at (-1.4, 2.0) {\HepParticle{\Pe}{R}{-}};
            \vertex (2) at (-1.4, 0.0) {\HepParticle{\Pe}{R}{+}};
            \vertex[dot,red] (A) at (0.0, 1.0) {};
            \vertex (3) at (1.4, 2.0) {\HepParticle{\PGm}{L}{-}};
            \vertex (4) at (1.4, 0.0) {\HepParticle{\PGm}{L}{+}};
            \diagram*{
                (1) -- [fermion] (A) -- [fermion] (2),
                (4) -- [fermion] (A) -- [fermion] (3)
            };
        \end{feynman}
        \node [above=-2ex] {$C^{V,LR}_{{\Pe\Pe\PGm\PGm}}$};
    \end{tikzpicture}
    \begin{tikzpicture}
        \begin{feynman}[large]
            \vertex (1) at (-1.4, 2.0) {\HepParticle{\Pe}{R}{-}};
            \vertex (2) at (-1.4, 0.0) {\HepParticle{\Pe}{R}{+}};
            \vertex[dot,red] (A) at (0.0, 1.0) {};
            \vertex (3) at (1.4, 2.0) {\HepParticle{\PGm}{L}{-}};
            \vertex (4) at (1.4, 0.0) {\HepParticle{\PGm}{L}{+}};
            \diagram*{
                (1) -- [fermion] (A) -- [fermion] (2),
                (4) -- [fermion] (A) -- [fermion] (3)
            };
        \end{feynman}
        \node [above=-2ex] {$C^{V,LR}_{{\PGm\PGm\Pe\Pe}}$};
    \end{tikzpicture}
    \begin{tikzpicture}
        \begin{feynman}[large]
            \vertex (1) at (-1.4, 2.0) {\HepParticle{\Pe}{R}{-}};
            \vertex (2) at (-1.4, 0.0) {\HepParticle{\Pe}{R}{+}};
            \vertex[dot,red] (A) at (0.0, 1.0) {};
            \vertex (3) at (1.4, 2.0) {\HepParticle{\PGm}{L}{-}};
            \vertex (4) at (1.4, 0.0) {\HepParticle{\PGm}{L}{+}};
            \diagram*{
                (1) -- [fermion] (A) -- [fermion] (2),
                (4) -- [fermion] (A) -- [fermion] (3)
            };
        \end{feynman}
        \node [above=-2ex] {$C^{V,LL}_{{\Pe\PGm\PGm\Pe}}$};
    \end{tikzpicture}
    \begin{tikzpicture}
        \begin{feynman}[large]
            \vertex (1) at (-1.4, 2.0) {\HepParticle{\Pe}{R}{-}};
            \vertex (2) at (-1.4, 0.0) {\HepParticle{\Pe}{R}{+}};
            \vertex[dot,red] (A) at (0.0, 1.0) {};
            \vertex (3) at (1.4, 2.0) {\HepParticle{\PGm}{L}{-}};
            \vertex (4) at (1.4, 0.0) {\HepParticle{\PGm}{L}{+}};
            \diagram*{
                (1) -- [fermion] (A) -- [fermion] (2),
                (4) -- [fermion] (A) -- [fermion] (3)
            };
        \end{feynman}
        \node [above=-2ex] {$C^{V,RR}_{{\Pe\PGm\PGm\Pe}}$};
    \end{tikzpicture}
    \begin{tikzpicture}
        \begin{feynman}[large]
            \vertex (1) at (-1.4, 2.0) {\HepParticle{\Pe}{R}{-}};
            \vertex (2) at (-1.4, 0.0) {\HepParticle{\Pe}{R}{+}};
            \vertex[dot,red] (A) at (0.0, 1.0) {};
            \vertex (3) at (1.4, 2.0) {\HepParticle{\PGm}{L}{-}};
            \vertex (4) at (1.4, 0.0) {\HepParticle{\PGm}{L}{+}};
            \diagram*{
                (1) -- [fermion] (A) -- [fermion] (2),
                (4) -- [fermion] (A) -- [fermion] (3)
            };
        \end{feynman}
        \node [above=-2ex] {$C^{V,LR}_{{\Pe\PGm\PGm\Pe}}$};
    \end{tikzpicture}
    \begin{tikzpicture}
        \begin{feynman}[large]
            \vertex (1) at (-1.4, 2.0) {\HepParticle{\Pe}{R}{-}};
            \vertex (2) at (-1.4, 0.0) {\HepParticle{\Pe}{R}{+}};
            \vertex[dot,red] (A) at (0.0, 1.0) {};
            \vertex (3) at (1.4, 2.0) {\HepParticle{\PGm}{L}{-}};
            \vertex (4) at (1.4, 0.0) {\HepParticle{\PGm}{L}{+}};
            \diagram*{
                (1) -- [fermion] (A) -- [fermion] (2),
                (4) -- [fermion] (A) -- [fermion] (3)
            };
        \end{feynman}
        \node [above=-2ex] {$C^{V,LR}_{{\PGm\Pe\Pe\PGm}}$};
    \end{tikzpicture}
    \begin{tikzpicture}
        \begin{feynman}[large]
            \vertex (1) at (-1.4, 2.0) {\HepParticle{\Pe}{R}{-}};
            \vertex (2) at (-1.4, 0.0) {\HepParticle{\Pe}{L}{+}};
            \vertex[dot,red] (A) at (0.0, 1.0) {};
            \vertex (3) at (1.4, 2.0) {\HepParticle{\PGm}{R}{-}};
            \vertex (4) at (1.4, 0.0) {\HepParticle{\PGm}{L}{+}};
            \diagram*{
                (1) -- [fermion] (A) -- [fermion] (2),
                (4) -- [fermion] (A) -- [fermion] (3)
            };
        \end{feynman}
        \node [above=-2ex] {$C^{S,RR}_{{\Pe\Pe\PGm\PGm}}$};
    \end{tikzpicture}
    \begin{tikzpicture}
        \begin{feynman}[large]
            \vertex (1) at (-1.4, 2.0) {\HepParticle{\Pe}{R}{-}};
            \vertex (2) at (-1.4, 0.0) {\HepParticle{\Pe}{L}{+}};
            \vertex[dot,red] (A) at (0.0, 1.0) {};
            \vertex (3) at (1.4, 2.0) {\HepParticle{\PGm}{R}{-}};
            \vertex (4) at (1.4, 0.0) {\HepParticle{\PGm}{L}{+}};
            \diagram*{
                (1) -- [fermion] (A) -- [fermion] (2),
                (4) -- [fermion] (A) -- [fermion] (3)
            };
        \end{feynman}
        \node [above=-2ex] {$C^{S,RR}_{{\Pe\PGm\PGm\Pe}}$};
    \end{tikzpicture}
    \caption{The 13 tree-level Feynman diagrams resulting from the LEFT operators and from QED.}
    \label{fig:LEFT_diag}
\end{figure}

These operators, along with QED, produce the 13 Feynman diagrams shown in figure~\ref{fig:LEFT_diag}.
In the limit of massless fermions, calculating the differential cross section from QED alone produces
\begin{equation*}
    \label{QED:diffXS}
    \frac{\mathrm{d} \sigma}{\mathrm{d}\cos \theta} = \frac{\pi \alpha^2}{2s} \left(1 + \cos^2 \theta\right),
\end{equation*}
and the inclusion of the LEFT diagrams produces, at leading order in LEFT (that is, up to order $1/\Lambda^2$),
\begin{align}
    \label{LEFT:diffXS}
    \begin{split}
        \frac{\mathrm{d} \sigma}{\mathrm{d} \cos \theta} = &
        \left[\frac{\pi \alpha^2}{2 s} + \frac{\alpha}{16} \frac{1}{\Lambda^2} \Re C_A \right]\left(1 + \cos^2 \theta\right) + \left[\frac{\alpha}{16} \frac{1}{\Lambda^2} \Re C_B \right] 2 \cos \theta,
    \end{split}
\end{align}
where $\alpha$ is the fine-structure constant, $\Lambda$ is the scale of new physics described by the LEFT, and $C_A$ and $C_B$ are linear combinations of Wilson coefficients,
\begin{align*}
    C_A &= C^{V,LL}_{\Pe\Pe\PGm\PGm} - C^{V,LL}_{\Pe\PGm\PGm\Pe} + C^{V,RR}_{\Pe\Pe\PGm\PGm} - C^{V,RR}_{\Pe\PGm\PGm\Pe} + C^{V,LR}_{\Pe\Pe\PGm\PGm} + C^{V,LR}_{\PGm\PGm\Pe\Pe} - 4 C^{S,RR}_{\Pe\PGm\PGm\Pe} \\
    C_B &= C^{V,LL}_{\Pe\Pe\PGm\PGm} - C^{V,LL}_{\Pe\PGm\PGm\Pe} + C^{V,RR}_{\Pe\Pe\PGm\PGm} - C^{V,RR}_{\Pe\PGm\PGm\Pe} - C^{V,LR}_{\Pe\Pe\PGm\PGm} - C^{V,LR}_{\PGm\PGm\Pe\Pe} + 4 C^{S,RR}_{\Pe\PGm\PGm\Pe}.
\end{align*}
Only two linear combinations of a total of seven Wilson coefficients, $(C_A + C_B)/2 = C^{V,LL}_{\Pe\Pe\PGm\PGm} - C^{V,LL}_{\Pe\PGm\PGm\Pe} + C^{V,RR}_{\Pe\Pe\PGm\PGm} - C^{V,RR}_{\Pe\PGm\PGm\Pe}$ and $(C_A - C_B)/2 = C^{V,LR}_{\Pe\Pe\PGm\PGm} + C^{V,LR}_{\PGm\PGm\Pe\Pe} - 4 C^{S,RR}_{\Pe\PGm\PGm\Pe}$, affect the differential cross section at order $1/\Lambda^2$.
The remaining five operators have effects beginning at order $1/\Lambda^4$.
Furthermore, $C_A$ affects the total cross section but not the forward-backward asymmetry, while $C_B$ affects the asymmetry but not the total cross section.

Although it is possible to calculate the quadratic contributions from these operators, this would not be the complete cross section at order $1/\Lambda^4$.
The complete cross section at order $1/\Lambda^4$ includes contributions from dimension-8 operators as well as contributions from Feynman diagrams involving two LEFT vertices.
Because the cross section at order $1/\Lambda^4$ would be incomplete without these contributions, we choose to truncate at order $1/\Lambda^2$, following established practice~\cite{LHCEFTWGvalidity}.

\section{LEFT fit results}
\label{sec:fit}

To measure the LEFT Wilson coefficients, we perform a Bayesian analysis.
We integrate eq.~\eqref{LEFT:diffXS} over each bin, multiply by $s$, and divide by the width of the bin,
\begin{equation*}
    \sigma^\text{exp.}_i\left(C_A, C_B\right) = \int_{D_i}^{U_i} \mathrm{d}\cos \theta \frac{s}{U_i - D_i} \frac{\mathrm{d}\sigma}{\mathrm{d}\cos \theta}(C_A, C_B),
\end{equation*}
where $\sigma^\text{exp.}_i$ is the predicted measurement in the $i$th bin, and $D_i$ and $U_i$ are respectively the lower and upper edges of the $i$th bin as shown in tables~\ref{tab:diffXS} and \ref{tab:diffXS2}.
We compare the measurement to the prediction using a Gaussian likelihood,
\begin{equation*}
    \mathcal{L}\left(C_A, C_B\right) = \prod_i \frac{1}{(\Delta \sigma^\text{obs.}_i) \sqrt{2 \pi}} e^{- \frac{1}{2} \left(\frac{\sigma^\text{exp.}_i - \sigma^\text{obs.}_i}{\Delta \sigma^\text{obs.}_i}\right)^2},
\end{equation*}
where $\sigma^\text{obs.}_i$ and $\Delta \sigma^\text{obs.}_i$ are respectively the measured cross section and its uncertainty in the $i$th bin.
For the parameters $C_A$ and $C_B$, we use flat prior probability distributions.
The QED prediction is the leading-order result exactly as shown in eq.~\eqref{LEFT:diffXS}.
No uncertainty in the QED prediction, nor in $\alpha$, is assumed.

\begin{figure}
    \centering
    \includegraphics{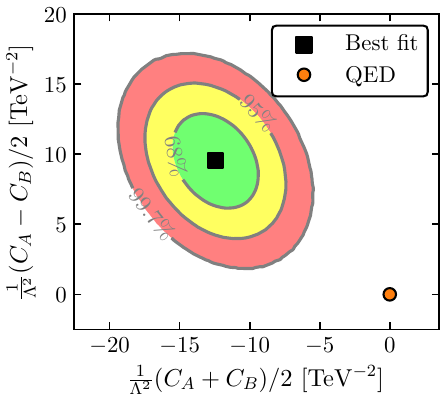}
    \caption{Posterior probability density for the LEFT Wilson coefficients.  The green, yellow, and red regions contain 68, 95, and 99.7\% of the posterior probability, respectively.  The black square shows the location of the maximum posterior probability density.  The red dot shows the values of the LEFT Wilson coefficients predicted by QED alone.  QED alone is very strongly disfavored.}
    \label{fig:2D_post_QED}
\end{figure}

\begin{figure}
    \centering
    \includegraphics{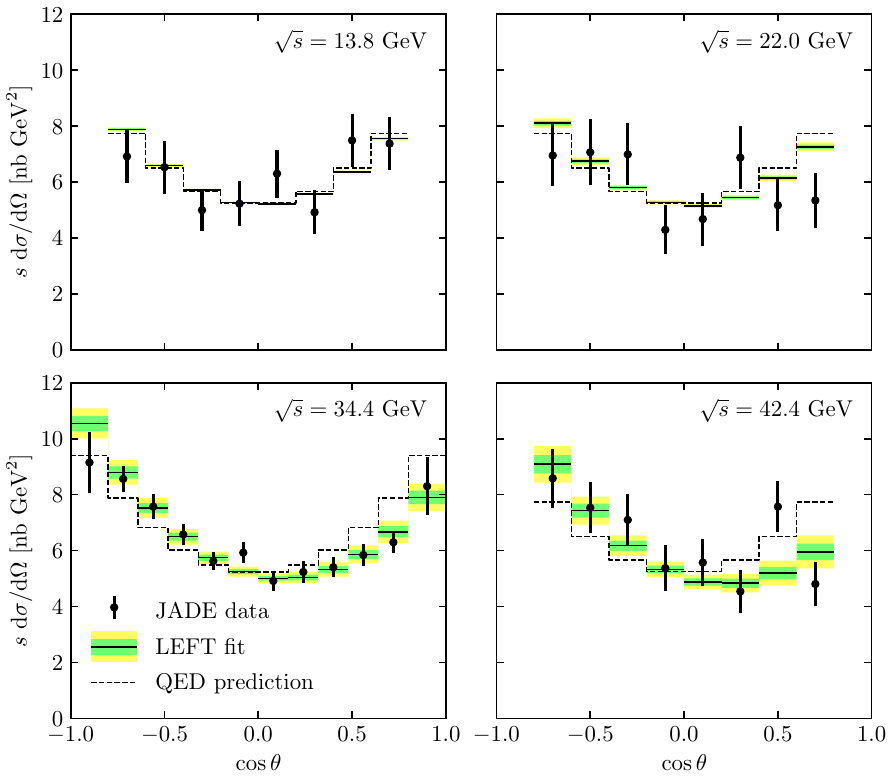}
    \caption{The JADE data (dots with error bars), compared to the prediction from QED alone (dashed line) and to the predictions resulting from the fit to the LEFT (solid line, with 68 and 95\% credible intervals shown in green and yellow, respectively).  The JADE data is inconsistent with QED alone, especially at higher center-of-mass energies, but it is consistent with the LEFT predictions.}
    \label{fig:JADE_noS_QED}
\end{figure}

Using the \textsc{pymc} software package~\cite{pymc5}, we draw samples from the posterior probability distribution.
From these samples, we construct the 68, 95, and 99.7\% highest-posterior-density credible regions, which are shown in figure~\ref{fig:2D_post_QED}.
We also compare the LEFT fit results and the predictions of QED alone to the JADE data, as shown in figure~\ref{fig:JADE_noS_QED}.

The prediction of QED alone, without any electroweak contributions, requires that the LEFT Wilson coefficients are all zero.
Figure~\ref{fig:2D_post_QED} shows that QED alone is very strongly disfavored.
In other words, from this JADE data, we have ``discovered'' physics beyond QED.

This is the situation in which we hope to find ourselves when measuring SMEFT Wilson coefficients at the LHC, that we measure some Wilson coefficients and strongly disfavor the SM.
The central question that is addressed by this case study is what happens next, and what this measurement can tell us about the new physics that we have observed.

\section{Matching to the electroweak theory}
\label{sec:match}

In the event that some Wilson coefficient measurement strongly disfavors the SM, one would look for models of physics beyond the SM, and ask what Wilson coefficients those models would predict as a function of the model parameters.
The effective field theory formalism allows many models to be directly compared to the data without requiring a dedicated measurement for each model.

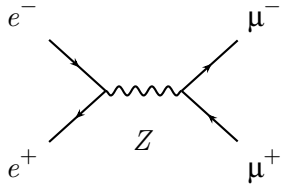
\begin{figure}
    \centering
    \begin{tikzpicture}
        \begin{feynman}[large]
            \vertex (1) at (-1.6, 2.0) {\HepParticle{\Pe}{}{-}};
            \vertex (2) at (-1.6, 0.0) {\HepParticle{\Pe}{}{+}};
            \vertex (A) at (-0.5, 1.0);
            \vertex (B) at (0.5, 1.0);
            \vertex (3) at (1.6, 2.0) {\HepParticle{\PGm}{}{-}};
            \vertex (4) at (1.6, 0.0) {\HepParticle{\PGm}{}{+}};
            \diagram*{
                (1) -- [fermion] (A) -- [fermion] (2),
                (4) -- [fermion] (B) -- [fermion] (3),
                (A) -- [yourboson] (B)
            };
        \end{feynman}
        \node [above] {\PZ};
    \end{tikzpicture}
    \caption{The tree-level \PZ boson exchange Feynman diagram from the electroweak theory.}
    \label{fig:EWK_diag}
\end{figure}

In the case of the JADE data, the obvious model to consider is the electroweak theory~\cite{Glashow:1961tr,Weinberg:1967tq,Salam:1968rm}.
At tree level, the electroweak theory adds one Feynman diagram, which contains a \PZ boson in the $s$ channel, as shown in figure~\ref{fig:EWK_diag}.
We can calculate the differential cross section of the electroweak theory in the limit of massless fermions and a zero-width \PZ boson,
\begin{align}
    \label{EWK:diffXS}
    \begin{split}
        \frac{\mathrm{d} \sigma}{\mathrm{d} \cos \theta} =& \frac{\pi \alpha^2}{2 s} \left(1 + \cos^2 \theta\right)\\
        &+ \frac{s}{\pi} \left(\frac{G_F M_\PZ^2}{s - M_\PZ^2}\right)^2 \left[\left(g_V^2 + g_A^2\right)^2 \left(1 + \cos^2 \theta\right) + 4 g_V^2 g_A^2 2 \cos \theta\right] \\
        &+ \sqrt{2} \alpha \frac{G_F M_\PZ^2}{s - M_\PZ^2} \left[g_V^2 \left(1 + \cos^2 \theta\right) + g_A^2 2 \cos \theta\right],
    \end{split}
\end{align}
where $G_F$ is Fermi's constant, $M_\PZ$ is the mass of the \PZ boson, $g_V = \sin^2 \theta_W - 1/4$ and $g_A = -1/4$ are the vector and axial-vector couplings of the \PZ boson to the electron and muon, and $\theta_W$ is the weak mixing angle.
Comparing eqs.~\ref{LEFT:diffXS} and \ref{EWK:diffXS} in the limit $s \ll M_Z^2$ (so that the EFT expansion parameter $s / \Lambda^2$ is small), we obtain the electroweak predictions for the LEFT Wilson coefficients,
\begin{align*}
    \Re \frac{C_A}{\Lambda^2} &= -16 \sqrt{2} G_F g_V^2\\
    \Re \frac{C_B}{\Lambda^2} &= -16 \sqrt{2} G_F g_A^2.
\end{align*}
Using the leading order relationship among $G_F$, $\sin^2 \theta_W$, $M_\PW$, and $M_\PZ$,
\begin{align*}
    M_\PW^2 &= \frac{\pi \alpha}{\sqrt{2} G_F \sin^2 \theta_W}\\
    M_\PZ^2 &= \frac{\pi \alpha}{\sqrt{2} G_F \left(1 - \sin^2 \theta_W\right) \sin^2 \theta_W},
\end{align*}
we can also express the Wilson coefficients as functions of $M_\PW$ and $M_\PZ$,
\begin{align}
    \label{eq:WCsfromMs}
    \begin{split}
        \Re \frac{C_A}{\Lambda^2} &= - \frac{16 \pi \alpha \left(1 - \frac{M_\PW^{2}}{M_\PZ^{2}} - \frac{1}{4}\right)^{2}}{M_\PW^{2} \left(1 - \frac{M_\PW^{2}}{M_\PZ^{2}}\right)}\\
        \Re \frac{C_B}{\Lambda^2} &= - \frac{\pi \alpha}{M_\PW^{2} \left(1 - \frac{M_\PW^{2}}{M_\PZ^{2}}\right)}.
    \end{split}
\end{align}

\section{Extracting the weak boson masses}
\label{sec:masses}

Now that we have predictions for the LEFT Wilson coefficients as functions of the parameters of the electroweak theory, we can reformulate the posterior probability density for the LEFT Wilson coefficients as a posterior probability density for the electroweak boson masses $M_\PW$ and $M_\PZ$.

\begin{figure}
    \centering
    \includegraphics{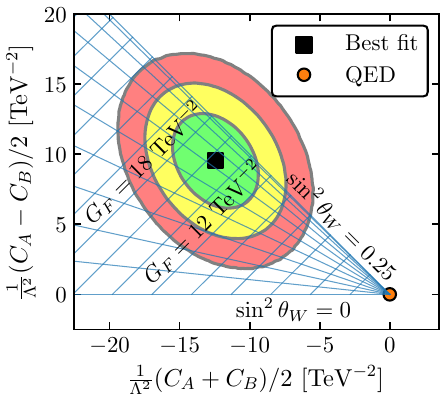}
    \caption{Posterior probability density for the LEFT Wilson coefficients, with contours of constant $G_F$ and contours of constant $\sin^2 \theta_W$ overlaid.}
    \label{fig:2D_post_overlay_noS}
\end{figure}

In figure~\ref{fig:2D_post_overlay_noS}, we overlay lines of constant $G_F$ and lines of constant $\sin^2 \theta_W$ on the posterior probability density of figure~\ref{fig:2D_post_QED}.
As $G_F$ approaches 0, we recover the QED-only prediction.
At $\sin^2 \theta_W = 0$, $C_A - C_B = 0$, and as $\sin^2 \theta_W$ approaches 0.25, we have $C_A = 0$.
As $\sin^2 \theta_W$ continues to increase beyond 0.25, we move back downwards in figure~\ref{fig:2D_post_overlay_noS}, so that the line for $\sin^2 \theta_W = 0.5$ lies on top of the line for $\sin^2 \theta_W = 0$.
This implies that the portion of the posterior probability density that lies above and to the right of the line for $\sin^2 \theta_W = 0.25$ is forbidden, and the rest of the space is double-covered.
In the main region of interest, $G_F$ is determined by $C_B$, which produces the forward-backward asymmetry, and $\sin^2 \theta_W$ is primarily determined by $C_A$, which only affects the total cross section.

To obtain the posterior probability density for $M_\PW$ and $M_\PZ$, we first calculate the values of $M_\PW$ and $M_\PZ$ that correspond to each sample from the posterior probability distribution over $C_A$ and $C_B$.
Because of the double covering, or equivalently because eq.~\eqref{eq:WCsfromMs} is quadratic in $M_\PW$ and $M_\PZ$, there are two $M_\PW$ and $M_\PZ$ solutions for each sample.
We calculate both solutions and treat them as two distinct samples.
Then, we reweight the samples to produce the posterior probability density that corresponds to a flat prior probability density over $M_\PW$ and $M_\PZ$ instead of a flat prior probability density over $C_A$ and $C_B$, using a weight for the $i$th sample
\begin{align*}
    w_i &= \frac{1}{\left\lvert J_i\right\rvert},
\end{align*}
where $\left\lvert J_i \right\rvert$ is the Jacobian determinant of eq.~\eqref{eq:WCsfromMs},
\begin{align*}
    \left\lvert J_i \right\rvert &= \begin{vmatrix}
        \frac{\partial \Re \frac{C_A}{\Lambda^2}}{\partial M_\PW} & \frac{\partial \Re \frac{C_A}{\Lambda^2}}{\partial M_\PZ} \\
        \frac{\partial \Re \frac{C_B}{\Lambda^2}}{\partial M_\PW} & \frac{\partial \Re \frac{C_B}{\Lambda^2}}{\partial M_\PZ}
    \end{vmatrix}_{\substack{M_\PW = M_{\PW}^{i}\\ M_\PZ = M_{\PZ}^{i}}}
    = \frac{\pi^{2} \alpha^{2} \left\lvert 128 {M_{\PW}^{i}}^{2} - 96 {M_{\PZ}^{i}}^{2}\right\rvert}{{M_{\PW}^{i}}^{3} M_{\PZ}^{i} \left({M_{\PZ}^{i}}^{2} - {M_{\PW}^{i}}^{2}\right)^2}.
\end{align*}
This procedure assigns zero prior probability to the samples that fall in the forbidden region, effectively removing them from consideration.

\begin{figure}
    \centering
    \includegraphics{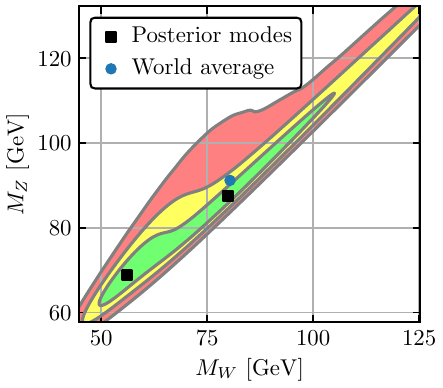}
    \includegraphics{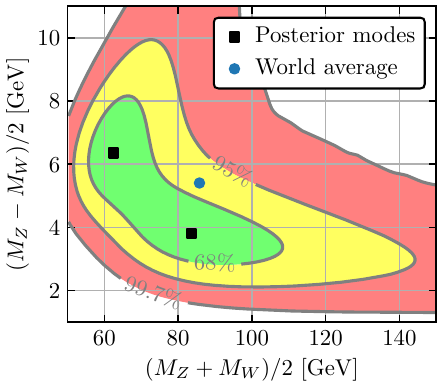}
    \caption{The measured masses of the \PW and \PZ bosons along with 68, 95, and 99.7\% credible regions, shown in green, yellow, and red, respectively.  The black squares show the local maxima of the posterior probability densities, and the blue dots show the current world average~\cite{PDG2024}.  The left plot shows the \PW and \PZ boson masses, while the right plot shows the average of the \PW and \PZ boson masses and half the difference between their masses.  This measurement agrees with the world average at the level of roughly $1.4\sigma$.}
    \label{fig:MW_MZ}
\end{figure}

The resulting posterior probability density over $M_\PW$ and $M_\PZ$ is shown in figure~\ref{fig:MW_MZ} along with the posterior probability density for $(M_\PW - M_\PZ)/2$ and $(M_\PW + M_\PZ)/2$.
The double covering results in a bimodal distribution.
The mode corresponding to $\sin^2 \theta_W < 0.25$ is at $M_\PW = \SI{79.9}{\giga\eV}$ and $M_\PZ = \SI{87.5}{\giga\eV}$ and the mode corresponding to $\sin^2 \theta_W > 0.25$ is at $M_\PW = \SI{56.1}{\giga\eV}$ and $M_\PZ = \SI{68.8}{\giga\eV}$.
The JADE data, considered through the lens of the LEFT, provide a measurement of the \PW and \PZ boson masses that is remarkably accurate, albeit with large uncertainties.
Comparing to the current world average~\cite{PDG2024}, we find that the level of agreement is roughly $1.4\sigma$, which is comparable to the roughly $1.3\sigma$ deviation between the forward-backward asymmetry measured by JADE and the electroweak prediction~\cite{JADE_mu_AFB}.

\begin{figure}
    \centering
    \includegraphics{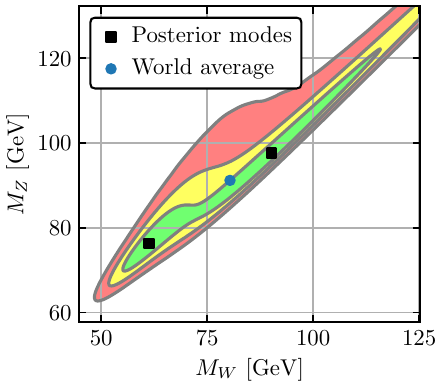}
    \includegraphics{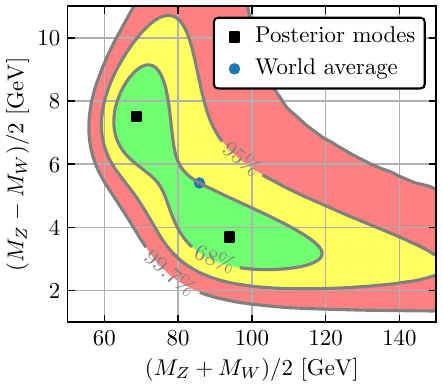}
    \caption{Results of a Bayesian analysis of the JADE data using the electroweak model, in the form of eq.~\eqref{EWK:diffXS} directly, instead of the LEFT.  Compared to the reinterpreted LEFT result, slightly higher boson masses are preferred.  This result is compatible with the LEFT result.}
    \label{fig:EWK_fit}
\end{figure}

As a cross-check, we perform a second Bayesian analysis using eq.~\eqref{EWK:diffXS} directly, imposing flat prior probability densities on $M_W$ and $M_Z$.
The results, which are compatible with the reinterpreted LEFT analysis, are shown in figure~\ref{fig:EWK_fit}.
Because eq.~\eqref{EWK:diffXS} includes the full $s$ dependence of the electroweak theory, the posterior probability density is shifted to slightly higher masses.
The overall compatibility demonstrates that it is not necessary to reanalyze data using a UV-complete model.
A reinterpretation of EFT results by matching to the UV-complete model is sufficient.

The remarkable result of measuring the \PW and \PZ boson masses through the lens of LEFT is obtained in spite of a number of possible deficiencies in the analysis.
Notably, at $\sqrt{s} = \SI{42.4}{\giga\eV}$, the EFT expansion parameter $s / \Lambda^2 \approx 1/2$ is not small.
Accordingly, we might anticipate that EFT contributions at order $1 / \Lambda^2$ might be insufficient, but this result appears to suggest that $1 / \Lambda^2$ is adequate even though $s$ is close to $\Lambda$.

There are a number of conceivable improvements that could be made to the analysis, including higher order corrections in the QED, electroweak, and LEFT calculations, and more careful handling of uncertainties in the JADE data.
These improvements have not been pursued, primarily because of a lack of information both about the corrections, ``up to order $\alpha^3$''~\cite{JADE_mu_AFB}, that were applied to the JADE measurement and about the details of the uncertainties in the JADE data.
Moreover, the level of agreement achieved between the result and our modern knowledge of the electroweak bosons suggests that these effects are negligible, and do not impact the fundamental conclusions about the utility of effective field theory.

This measurement of the \PW and \PZ boson masses would have been sufficient to guide the construction of then-future colliders such as the super proton-antiproton synchrotron, which discovered the \PW and \PZ bosons~\cite{UA1:1983crd,UA1:1983mne,UA2:1983mlz,UA2:1983tsx}, and the large electron-positron collider, which measured the properties of the \PZ boson in unsurpassed detail~\cite{ALEPH:2005ab}.
Accordingly, we anticipate that, given an observation of SMEFT Wilson coefficients in significant tension with the SM at the LHC, matching to UV-complete models will permit sufficient understanding of the parameters of those models to guide the construction of future colliders such as ILC, CLiC, FCC-ee, FCC-hh, CEPC, or a muon collider.

\section{Conclusion}
\label{sec:conclusion}

The low-energy effective field theory provides an adequate description of the JADE $\Pep\Pem \to \PGmp\PGmm$ data below the \PZ boson mass.
It permits the observation of physics beyond QED with a high level of significance, more than 5 standard deviations.
Furthermore, by matching the measured Wilson coefficients of the low-energy effective field theory to the electroweak theory, we can obtain a rough measurement of the masses of the \PW and \PZ bosons.
This measurement would have been sufficient, even in the absence of other data, to guide the construction of the super proton-antiproton synchrotron and the large electron-positron collider.

Accordingly, as we search for signs of physics beyond the standard model using the standard model effective field theory, we anticipate that a discovery will provide sufficient information, by matching to one or more UV-complete models, to guide the construction of future colliders such as ILC, CLiC, FCC-ee, FCC-hh, CEPC, or a muon collider.
This case study demonstrates both the limitations and the power of effective field theory as a tool to discover and characterize new physics, and provides hope and guidance to the effective field theory efforts at the LHC and beyond.

\acknowledgments

We would like to thank Jennet Dickinson, Michael Peskin, and Dante Amidei for helpful and inspiring discussions and suggestions.
This work was supported by DOE grant DE-SC0007861.

\bibliographystyle{JHEP}
\bibliography{biblio.bib}

\end{document}